\documentclass[iop]{amsart}
\usepackage{amssymb}
\usepackage{amsthm}
\usepackage[english]{babel}
\usepackage[T1]{fontenc}
\usepackage[cp1250]{inputenc}
\usepackage{longtable}
\usepackage[a4paper,outer=3cm,inner=3cm]{geometry}
\usepackage{fancyhdr}
\usepackage{rotating}

        \title{Parabolic Strip Telescope}
        \author{G.Chadzitaskos
\\ Department of Physics,\\
Faculty of Nuclear Sciences and Physical Engineering,\\ Czech
Technical University, B\v rehov\'a 7, CZ - 115 19 Prague\\e-mail:
goce.chadzitaskos@fjfi.cvut.cz}
       
\begin{document}
        \maketitle
 \begin{abstract}
  We present a proposal of a new type of telescopes using a rotating parabolic strip as
  the primary mirror. It is the most principal modification of the design of telescopes from the times of
  Galileo and Newton. In order to demonstrate the basic idea, the image of an artificial
  constellation observed by this kind of telescope was reconstructed using the techniques described in this article.
  As a working model of this new telescope, we have used an assembly of
  the primary mirror---a strip of acrylic glass parabolic mirror 40 cm long and 10 cm wide
  shaped as a parabolic cylinder of focal length 1 m---and an artificial
  constellation, a set of 5 apertures in a distance of 5 m illuminated from behind. In order  to reconstruct the image, we made
  a series of snaps, each after a rotation of the constellation by 15 degrees. Using Matlab we reconstructed the image of
  the artificial constellation.

 \end{abstract}


\section{Introduction}

All known telescopic systems are of two basic types: the
Galileo--Kepler  using lenses (or lenses with corrections of optical
defects) as primary optical elements (refractors),  or  the
Newton--Cassegrain  with  spherical or paraboloid mirrors
(reflectors) as the primary elements.  The image is observed using
photographic cameras, CCD cameras or a light spectrometers. The
images from the CCD or the spectrometers are stored in computers for
further elaboration. In all cases, the images are exposed in the
image plane. The primary elements of the reflectors and of the
refractors collect parallel rays of light in the focal plane, and
other optical elements are used in different geometrical
configurations. For more details see \cite{a}.

As a rule,  telescopes are supported by  mountings which make it
possible to point their optical axes to the object to be observed.
In the case of moving objects, the mounting makes the optical axis
able to follow the position of the object. There are several types
of mountings, the most important are  parallactic, azimuthal, and
four axes mountings. The azimuthal mounting has horizontal and
vertical rotational axes, and it is necessary to perform both
rotations simultaneously during the observation of moving objects.
The parallactic mounting has two perpendicular axes, one axis is
parallel to the Earth's axis, and a suitable rotation around this
axis makes the telescope follow the celestial objects. Most
contemporary  mountings are controlled by computers.

Up to now, telescopes have their primary elements of circular or
regular polygonal form. The angular resolution of such telescopes is
almost the same in all lateral directions, and it is defined as the
smallest angle between two close points which can be distinguished.
The angular resolution is the ability of a telescope to render
detail: the higher resolution, the finer details can be observed.
The angular resolution, together with the aperture size, are the two
most important characteristics of telescopes. In the case of
circular primary mirror of diameter $D$, its area $P$ and angular
resolution $\delta$ are
$$ P = \pi \frac{D^{2}}{4},\,\,\,\,\, \delta \approx 1,22 \frac{\lambda}{D} $$
for a monochromatic light of wavelength $ \lambda $ \cite{b}.

 There are two reasons for the
construction of very large telescopes: to gain better angular
resolution and to collect more light \cite{c}. In those cases
improving the resolution and collecting more light means increasing
the diameter of the primary mirror.

\section{Telescope with a rotating objective element}

The basic idea of our new system has been inspired by X-ray computer
tomography (CT). An X-ray source is located on one side and
an X-ray camera on the opposite side of a sample. The integral
absorptions of X-rays in different angles step by step during the
rotation of the sample are measured. The total absorption of all photons
coming along different lines perpendicular to the camera are
registered as points of a one-dimensional picture. Finally, the
inverse Radon transform is used to reconstruct the image of
absorption of X-rays in different points of the media.

Modern mathematical methods and software developed for CT involve
gathering projection data from multiple directions and feeding the
data into a tomographic reconstruction software algorithm processed
by a computer \cite{c}. Mathematical filters are used to improve the
reconstructed image \cite{d}. The idea and the image processing of
the technology of Single-Photon Emission Computer Tomography
(English 1996) is the same as for our parabolic strip telescope.
This technology is used in nuclear medicine where a patient is
injected with a radiopharmaceutic which emits gamma rays. The
emitted gamma rays are collected by a gamma camera and the emitted
image is reconstructed \cite{g}.

\begin{figure}[h]
      \includegraphics{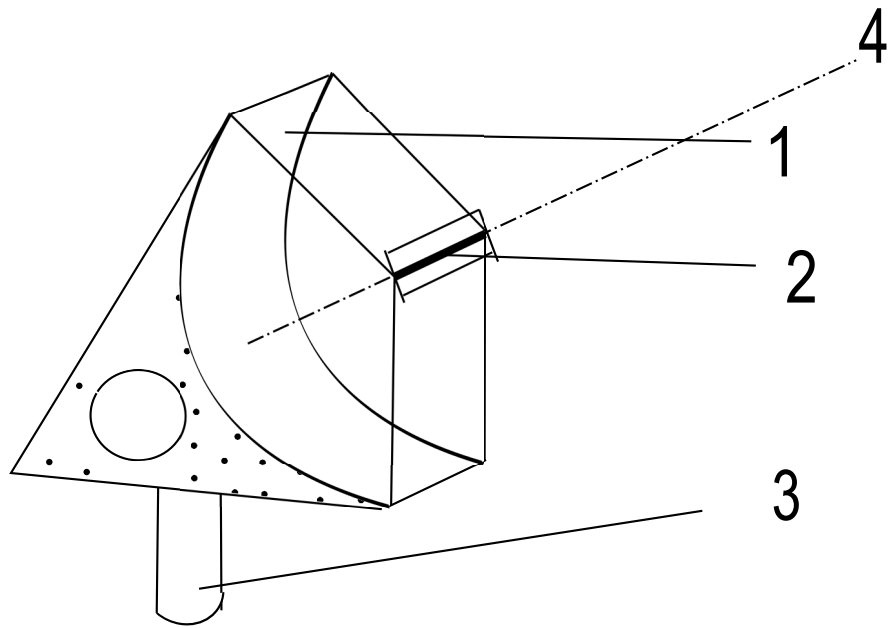}
      \caption{Principle of the new telescopic system. It consists of a
parabolic strip mirror 1,  CCD camera 2 in the image plane,
supported by mounting 3, and rotates around the optical axis 4. The
shots are stored in a computer, where the image is reconstructed. }
\label{fig:1}
 \end{figure}

The same approach can be used when a parabolic strip is the primary
mirror (h, i) of a telescope, following the schematic shown in
Figure \ref{fig:1}. The images of observed objects are lines. Each
line represents the integral intensity of light incoming from the
object or objects perpendicular to the strip (parallel to the focal
line) located inside the field of view  which is guaranteed by the
geometry. Making a series of photos while rotating the telescope
around its optical axis, one can use the inverse Radon transform to
reconstruct the image with the above mentioned angular resolution.
It is also possible to use other tools, for example Matlab \cite{g}.
Secondary optic elements can be used  to focus the lines from the
focal plane into points.

\begin{figure}[h]
      \includegraphics{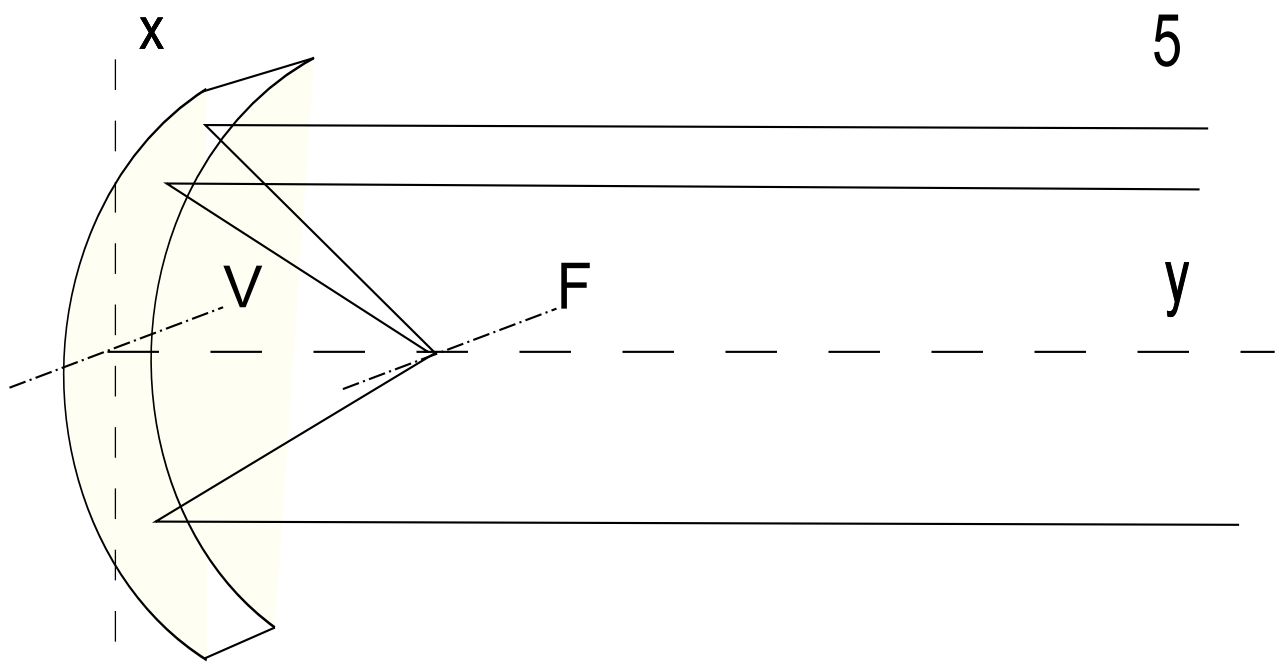}
      \caption{The paraxial beams 5 are reflected by strip on the focus line F, where V is the vertex line of the strip.}
      \label{fig:2}
 \end{figure}

The technological construction of the parabolic strip is simpler
than planar or paraboloid surface, because the stress of material
helps to maintain the geometry. A precise parabolic bracket is
required to hold the parabolic strip made of elastic mirrors. It is
also simpler to use adaptive optics for the correction of defects.
One can control the surface by a laser pointer located on one side
of the strip. By the detection of its reflected light on the other
side of the strip during the scan, the whole surface can be
controlled and corrected by corrective elements. Of course,
secondary mirrors can be added, then we have a Newton--like or
Cassegrain--like telescope. The area and the angular resolution of a
parabolic strip telescope are

$$ P=L W, \,\,\,\,\,\,\delta_{L} = \frac{\lambda}{L}, $$

where $\lambda$ is the wavelength of light, $L$ is the length of
projection of strip on tangent plane at the vertex line---in Figure
\ref{fig:2} it is the length of x-axis projection of the strip---and
$W$ is the width of the strip.

 The proposed telescopic system involves preferably parallactic mounting
 and an instrument for the digitalization of the image connected  to a computer.

 The mounting, however, has to perform one more rotation than the usual mounting:
 a rotation of the primary element or of the whole system around the optical axis is necessary, in order to
 reconstruct the details of the object.  The instrument for the digitalization of images has to be located in the
 image plane of the telescope.

For refractors the new primary elements are cuts of parabolic
cylinders and instruments for the digitalization of the image are
located in the image plane of the telescopes. Of course, this idea
can also be used for radio telescopes.

\section{Advantages}

  We have proposed a fundamental modification of reflectors where the angular
  resolution can be better than in the case of usual objectives of the same area. The
reconstruction of images exploits the discrete inverse Radon
transform \cite{f} or similar mathematical methods. The main
advantages of such telescopes are

\begin{itemize}
\item  good angular resolution,
\item  low expenses,
\item  simple technological development,
\item  possibility to install a grid of large telescopes across the Earth,
\item  lower weight for use on satellites.
\end{itemize}

The only major complication is that one more rotational movement is
needed to reconstruct the image with the same angular resolution in
all directions.

The good angular resolution can be used for direct observation of
bright objects. The additive rotation is not necessary for some
purpose, for example when the plane of rotation of rotating objects
is known, and changes of orbits are observed. For example, the
Jupiter-like planet around close stars can be directly detectable by
a 60 meter telescope, from point of view of the requirement angular
resolution.

The relatively low cost of parabolic strip telescopes makes it possible to
cover the Earth from North to South by a grid of telescopes. Each of
them would observe, for example, 30 degrees around the zenith. This would
reduce the cost of mountings and multiply the observation time.
The cost of telescopes observing $\pm$ 15 degrees will be reduced
because of:

\begin{itemize}
\item  high mounting not being necessary,
\item  low wind influence,
\item  low gravitational influence.
\end{itemize}

\section{Secondary optics}
The parabolic strip telescope offers the possibility to use
secondary optical elements to focus light from the primary mirror to
short segments or   points. But these secondary elements may
introduce errors or noise in resulting  images. The most natural and
easiest way is to use a "software" secondary element, i.e. a
computer process on images created by the primary mirror. This can
be explained on the following simple example. Let us observe two
distinguishable objects lying on a line parallel to the vertex line.
The points represent objects whose structure is below the angular
resolution of a telescope in that direction. The image of each
object is a segment of length W that is parallel with the vertex
line, and both segments lie on the same straight line. The centers
of the segments are located symmetrically to the optical axis in
directions according to the reflection law. Depending on the strip
width W and the angle between the two objects, the two segments are
either separated or overlapping. If the segments are separated,
their images are placed in the centers of  the segments. If  the
segments are overlapping, then a simple analysis  gives the position
of images of  the objects.  One has to subtract the half width of
the parabolic strip $\frac{W}{2}$ from one end of the overlapping
segments and  from the opposite end of the overlapping segments, and
these are the positions of objects on the image. In Figure 5 the end
points of displaced segments are shown. The segments are images of
the artificial constellation.
   Angular resolution in determining the position of a point in the direction of the vertex
   line is $\delta_{W}=\frac{ \lambda}{W}$ and is substantially less than in the perpendicular
   direction  $\delta_{L}=\frac{ \lambda}{L}$. This difference in angular resolution in two
   perpendicular directions can serve for observation of some particular phenomena. For example,
   to monitor fine movements of an object  in the known direction, the telescope has to be adjusted
   so that in this direction the best angular resolution is achieved.
   During  computer processing all the photons reflected from the primary mirror  are recorded
   by pixels of  a  CCD camera in the image plane and no one is lost.
   Using a secondary parabolic strip or a secondary parabolic lens
   from the Hartley disc, the segments can be focused into points. Then the inverse Radon transform reconstruct
   the image with appropriate angular resolutions.

\section{The proof-of-principle experiment}

\begin{figure}[h]
      \includegraphics{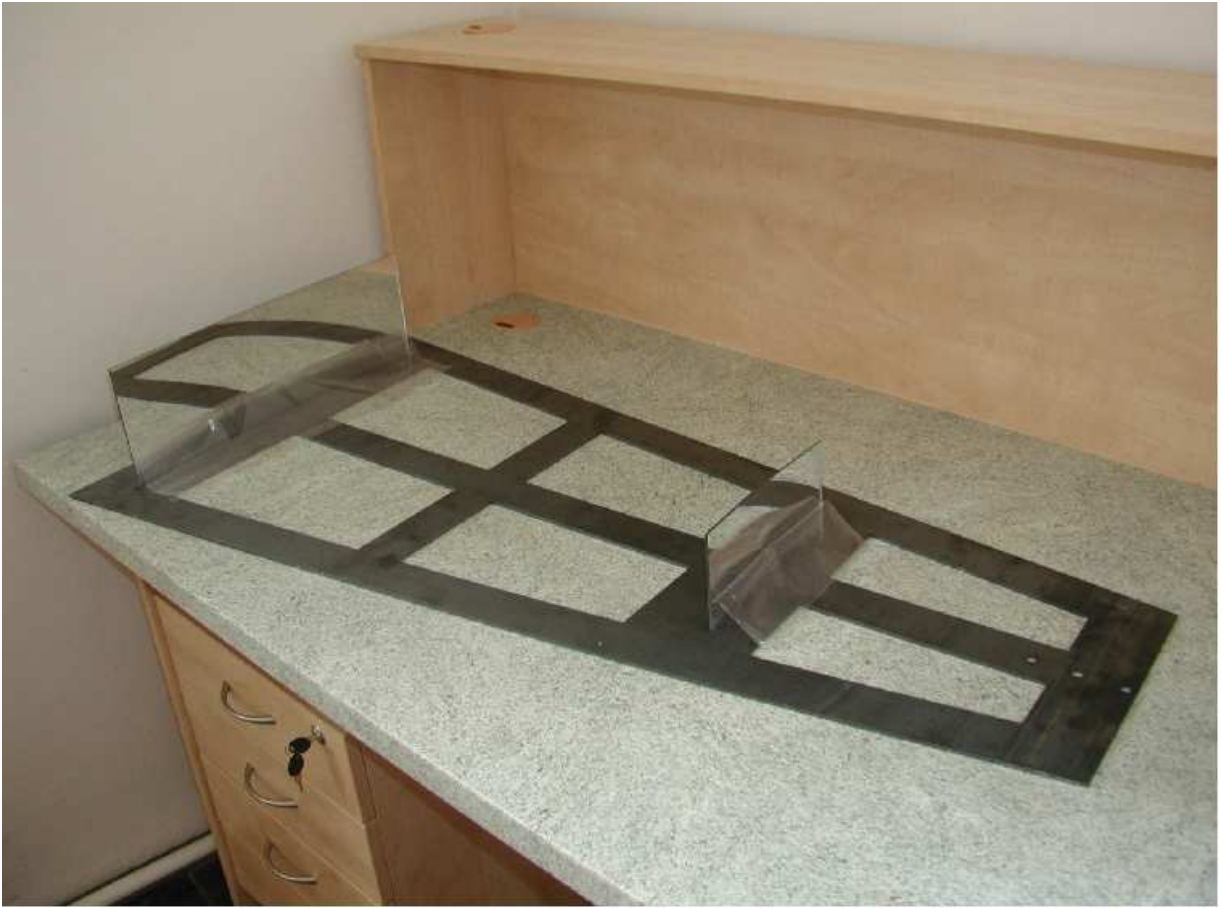}
      \caption{Two parabolic strips of lengths 20 and 40 cm, for the proof-of-principle experiment, the 40
cm strip was used}\label{fig:3}
 \end{figure}

\begin{figure}[h]
      \includegraphics{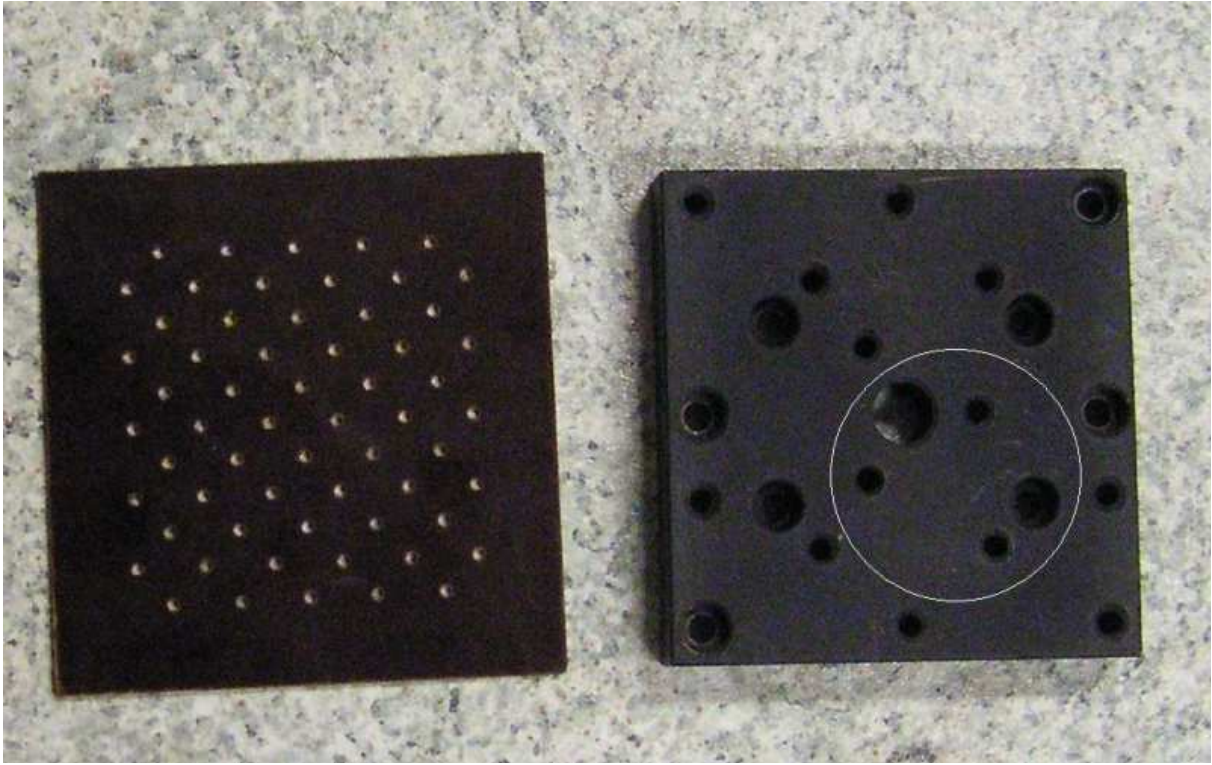}
      \caption{Two artificial constellations were used to demonstrate the principle. For presentation we use the
      second one. The circle encloses the illuminated area.}
      \label{fig:4}
 \end{figure}

\begin{figure}[h]
       \includegraphics{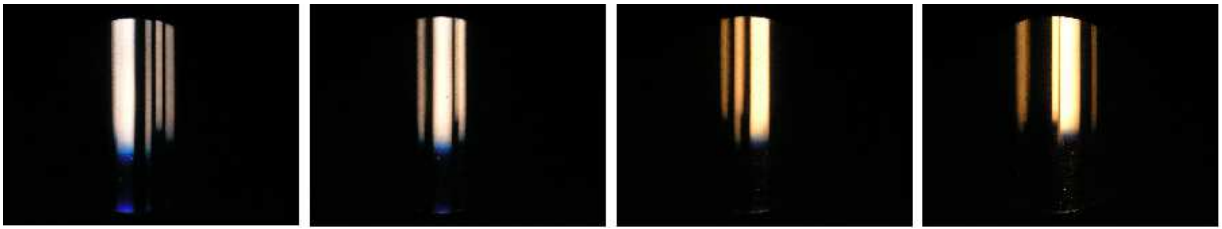}
       \caption{The original images from camera made at angels 0, 45, 90, and
135 degrees} \label{fig:5}
   \end{figure}

\begin{figure}[h]
        \includegraphics{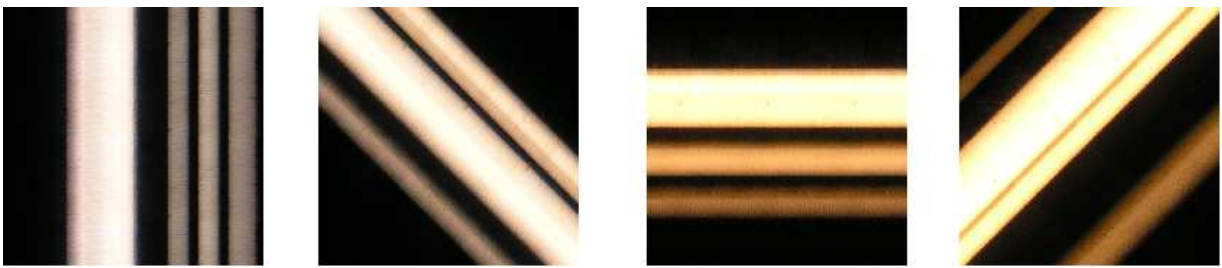}
        \caption{The same images after Matlab processing at corresponding
angles} \label{fig:6}
\end{figure}

\begin{figure}[h]
        \includegraphics{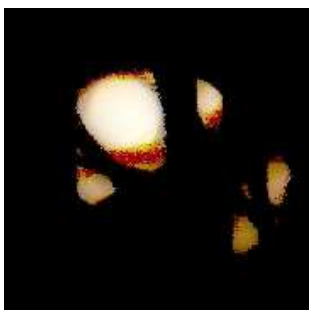}
        \caption{The image reconstructed from series of pictures made with angular step 15 degrees}
        \label{fig:7}
\end{figure}

 In order to show that the principle works we have prepared a very basic experiment.
  For the sake of simplification, the
 telescope was stationary. Figure \ref{fig:3} shows manufactured
 parabolic strips of lengths 20 and 40 cm. In the experiment the 40 cm strip was used.
  The artificial constellation was
 represented by a light shining through  five apertures.  The artificial
 constellations used are shown in Figure \ref{fig:4}.
 The constellation was then rotated in steps of 15 degrees and the
 images were photographed by an ordinary digital camera. Figure \ref{fig:5} shows four photos
 for the constellation at angles 0, 45, 90 and 135 degrees.  Figure
 \ref{fig:6}  shows the same photos after two steps of processing---averaging and rotating.
 The reconstructed image is shown on Figure \ref{fig:7}. For image
 processing, Matlab was used.

  The principle of the telescope was successfully tested. The resulting image can be compared with the reconstructed images in
  \cite{g}. With higher quality components and more measurements at finer angle steps, the reconstructed image would be
    of better quality.

\section*{Acknowledgements}
The support by the Ministry of Education of Czech Republic project
MSM6840770039  is acknowledged.

\end{document}